\documentclass[aps,pra,preprint]{revtex4}
\usepackage{graphicx,hyperref}
\usepackage{amsmath,amssymb}
\newcommand{\heff}{H_{\rm\scriptscriptstyle eff}}

\newcommand{\trl}{{\rm Tr}_{\rm\scriptscriptstyle E}}
\newcommand{\rhos}{\rho_{\rm\scriptscriptstyle S}}
\newcommand{\rhol}{\rho_{\rm\scriptscriptstyle E}^{\rm\scriptscriptstyle eq}}

\newcommand{\wsl}{\omega_{\rm\scriptscriptstyle SE}}

\newcommand{\tauc}{\tau_c}
\newcommand{\wone}{\omega_1}
\newcommand{\stauc}{\tau_c^{(s)}}
\newcommand{\swone}{\omega_1^{(s)}}

\begin{document}

\title{Optimal fidelity in implementing Grover's search algorithm on open quantum system}
\author{Nilanjana Chanda}
\email{nc16ip020@iiserkol.ac.in}
\author{Rangeet Bhattacharyya}
\email{rangeet@iiserkol.ac.in}
\affiliation{Department of Physical Sciences, Indian Institute of Science Education and Research Kolkata,
Mohanpur -- 741246, West Bengal, India}

\begin{abstract}

We investigate the fidelity of Grover's search algorithm by implementing it on an open quantum system. 
In particular, we study with what accuracy one can estimate that the algorithm would deliver the searched state.
In reality, every system has some influence of its environment. We include the environmental effects on the system dynamics by using a recently reported fluctuation-regulated quantum master equation (FRQME). The FRQME indicates that in addition to the regular relaxation due to system-environment coupling, the applied drive also causes dissipation in the system dynamics. 
As a result, the fidelity is found to depend on both the drive-induced dissipative terms and the relaxation terms and we find that there exists a competition between them, leading to an optimum value of the drive amplitude for which the fidelity becomes maximum.
For efficient implementation of the search algorithm, precise knowledge of this optimum drive amplitude is essential.


\end{abstract}

\maketitle

\section{Introduction}

Grover's search algorithm is used to search a certain element from an unstructured set of data \cite{grover_97}. 
It has been demonstrated theoretically and experimentally on two-qubit system long back \cite{chuang_98,jones_98}.
In this work, we use fluctuation-regulated quantum master equation (FRQME) \cite{chakrabarti_pra_2018} 
to study the second order effects of the applied drive on the fidelity of Grover's search algorithm, considering the composite system as an open quantum system. 
We determine the condition to achieve the optimal fidelity with which the search algorithm would produce the target state.

\section{Method}

To construct the FRQME,
we consider a driven quantum system coupled to its environment that undergoes thermal fluctuations. 
Using Born-Markov approximation and time coarse-graining procedure, the following dynamical equation for the system density matrix has been derived under an ensemble average,
\begin{eqnarray}
\label{eq1}\frac{d}{dt}{\rhos}(t) = &-& i \; \trl [\heff (t),{\rhos}(t)\otimes \rhol]^{\rm sec} 
\nonumber \\
&-& \int_0^\infty d\tau\; \trl [\heff(t),[\heff (t-\tau),{\rhos}(t)\otimes \rhol]]^{\rm sec}\; 
e^{-\frac{|\tau|}{\tau_c}}
\end{eqnarray}
where $\heff$ contains the drive as well as the system-environment coupling Hamiltonians, 
and 
$\tau_c$ is the characteristic timescale of decay of autocorrelation of environmental fluctuations. 
We note that the drive Hamiltonian appearing in the second order term causes dissipation in the system dynamics and is referred to as the {\it drive-induced dissipation} (DID) \cite{chakrabarti_pra_2018, chakrabarti_epl_2018}.

The FRQME (\ref{eq1}) can be expressed in Liouville space as,
\begin{equation}
\label{eqLiouville}\frac{d\hat{\rhos}}{dt} = 
[-i \hat{\hat{\mathcal{L}}}^{(1)}_{\rm drive} - \hat{\hat{\mathcal{L}}}^{(2)}_{\rm drive} 
- \hat{\hat{\mathcal{L}}}^{(2)}_{\rm system-env.}]\hat{\rhos}  
= \hat{\hat{\Gamma}} \hat{\rhos}.
\end{equation}

The solution of FRQME (\ref{eqLiouville}) is, 
\begin{equation}
\label{eq3} \hat{\rhos}^{\rm fin} = e^{\hat{\hat{\Gamma}} t} \hat{\rhos}^{\rm ini} = \mathcal{U} \hat{\rhos}^{\rm ini}.
\end{equation}

To study the effect of decoherence on Grover's search algorithm, we apply the FRQME on the qubit system. 
We implement the conditional sign-flip operation of the algorithm, considering that the initial state is prepared at a uniform superposition of all four basis states \{$|00\rangle, |01\rangle, |10\rangle, |11\rangle$\}.
According to the unitary dynamics, for the search of state $|01\rangle$, the expected state after the operation would be,
$ |\phi_1\rangle = \frac{1}{2} \begin{pmatrix}
1\\-1\\1\\1
\end{pmatrix} $.

We evaluate how the final state gets modified in the presence of the decoherence processes.

\section{Results and Discussions}

\subsection{Analytical}

At first, we incorporate the second order contribution coming from the drive only (through $\hat{\hat{\mathcal{L}}}^{(2)}_{\rm drive} $) to get a compact form analytical expression for fidelity.

For the final state $\rhos^{\rm fin}$, 
\begin{equation}
{\rm Tr}[\rhos^{\rm fin}]^2 = \frac{1}{64}  (17 + 2 e^{-18 \pi \omega_1 \tau_c} + 4 e^{-14 \pi \omega_1 \tau_c} + 3 e^{-10 \pi \omega_1 \tau_c} + 33 e^{- 8 \pi \omega_1 \tau_c} - 2 e^{-6 \pi \omega_1 \tau_c} - 2 e^{-4 \pi \omega_1 \tau_c} + 9 e^{-2 \pi \omega_1 \tau_c} ).
\end{equation}
For non-zero $\omega_1$ and $ \tau_c$, ${\rm Tr}[\rhos^{\rm fin}]^2 < 1 $. This implies that the final state is a mixed state and at the end of the operation, along with the target state $|\phi_1\rangle$, there is finite probability of getting other states as well.

\begin{figure}[h!]
\center
\includegraphics[width=0.6\linewidth]{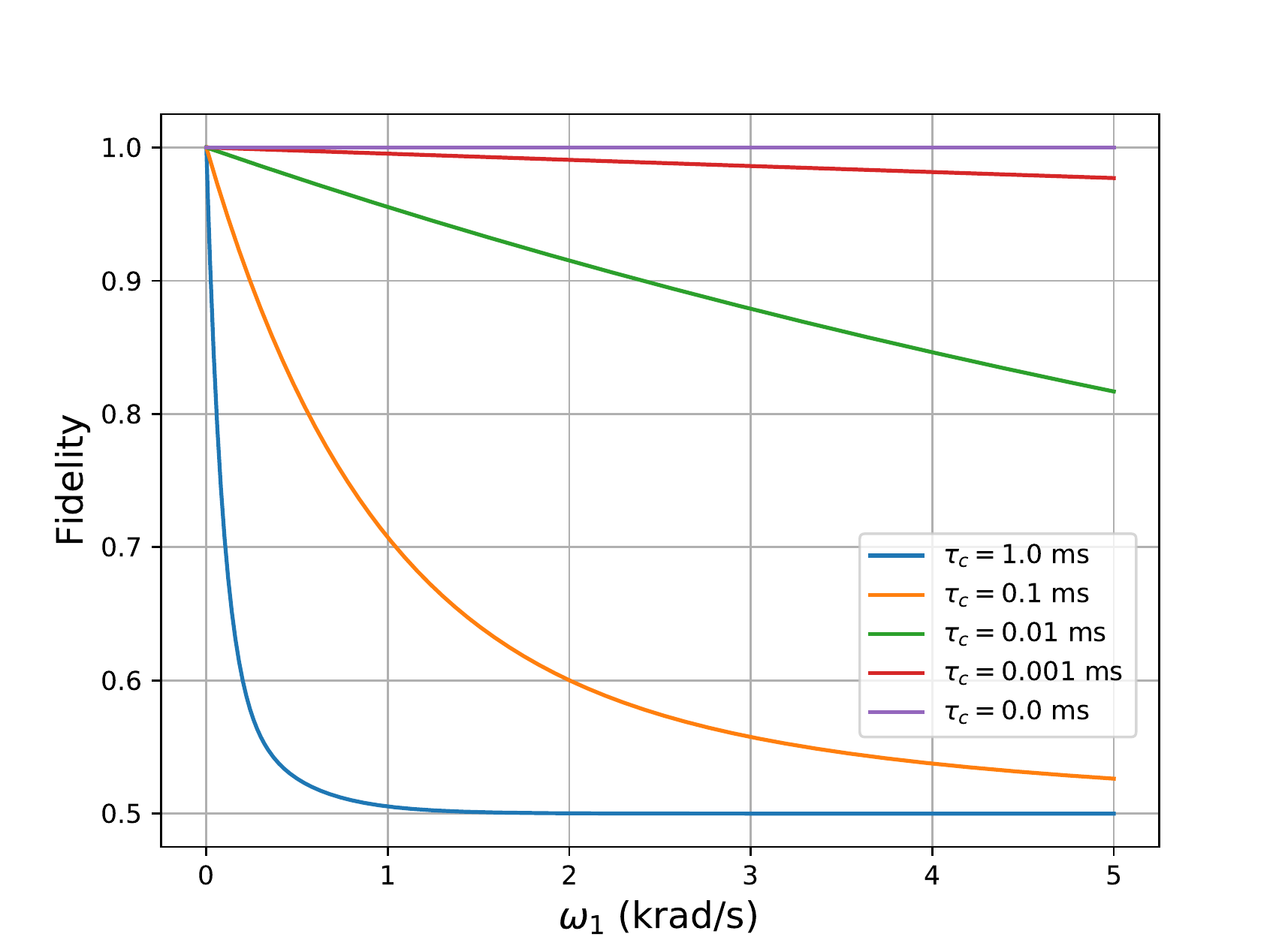}
\caption{Fidelity $F(|\phi_1\rangle, \rhos^{\rm fin})$ 
is plotted as a function of $\omega_1$ for different $\tau_c$ values, including only DID terms.}
\label{f1} 
\end{figure}

The fidelity between $|\phi_1\rangle$ and the obtained mixed state $\rhos^{\rm fin}$ is found to be, 
\begin{equation}
F(|\phi_1\rangle, \rhos^{\rm fin}) = \sqrt{\langle \phi_1| \rhos^{\rm fin} |\phi_1\rangle} %= \sqrt{p_1} 
= \frac{1}{4}  \sqrt{(4 + e^{-9 \pi \omega_1 \tau_c} + e^{-5 \pi \omega_1 \tau_c} + 8 e^{-4 \pi \omega_1 \tau_c} + 2 e^{- \pi \omega_1 \tau_c})}.
\end{equation}

When $\tau_c = 0$ (unitary case), the fidelity becomes $1$. 
As $\tau_c$ attains some non-zero value, the fidelity starts to decrease with $\omega_1$ due to DID and gradually approaches the value $0.5$, as can be verified from figure \ref{f1}.

\subsection{Numerical}

\begin{figure}[h!]
\center
\raisebox{40mm}{(a)}
\includegraphics[width=0.4\linewidth]{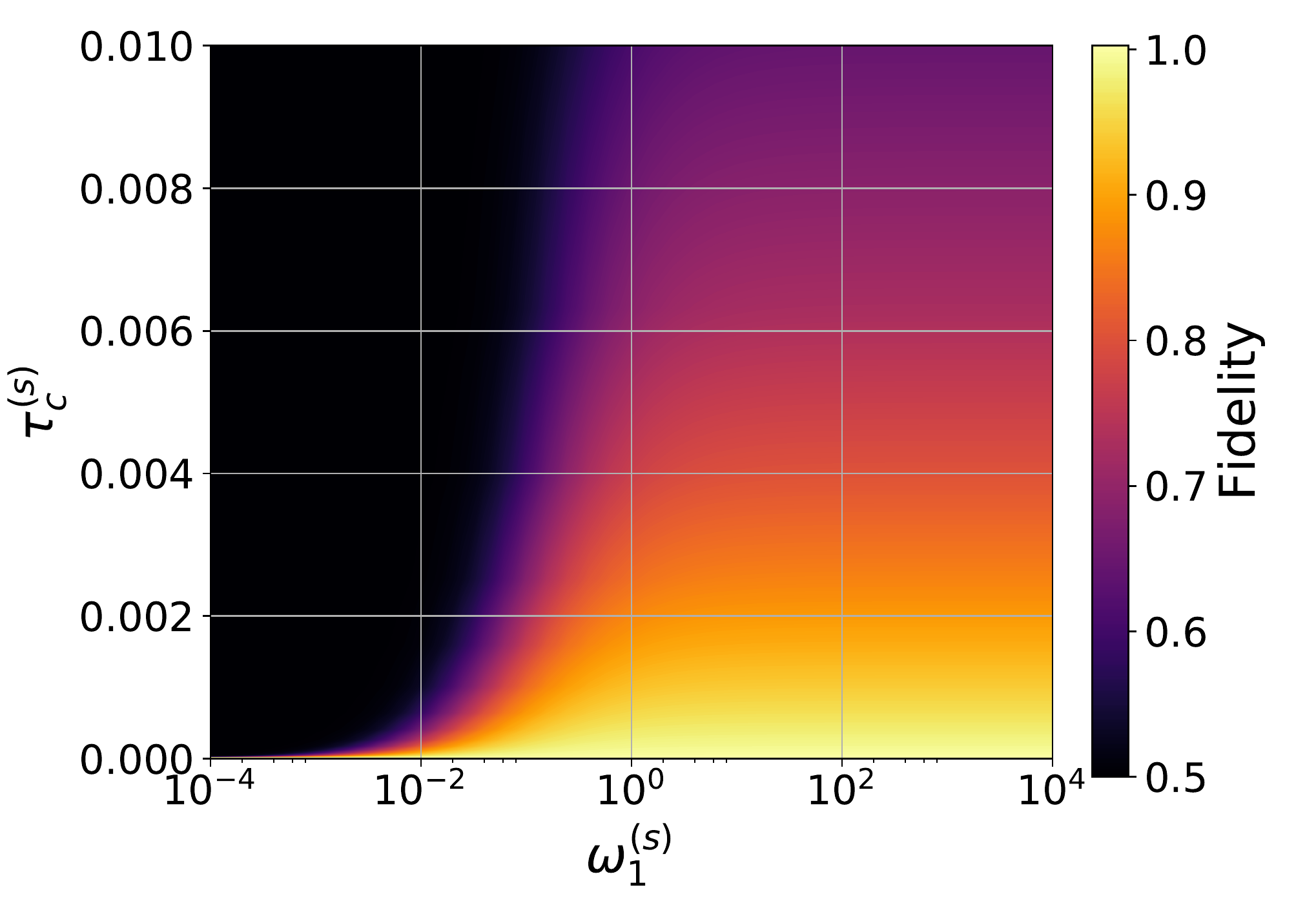}
\raisebox{40mm}{(b)}
\includegraphics[width=0.4\linewidth]{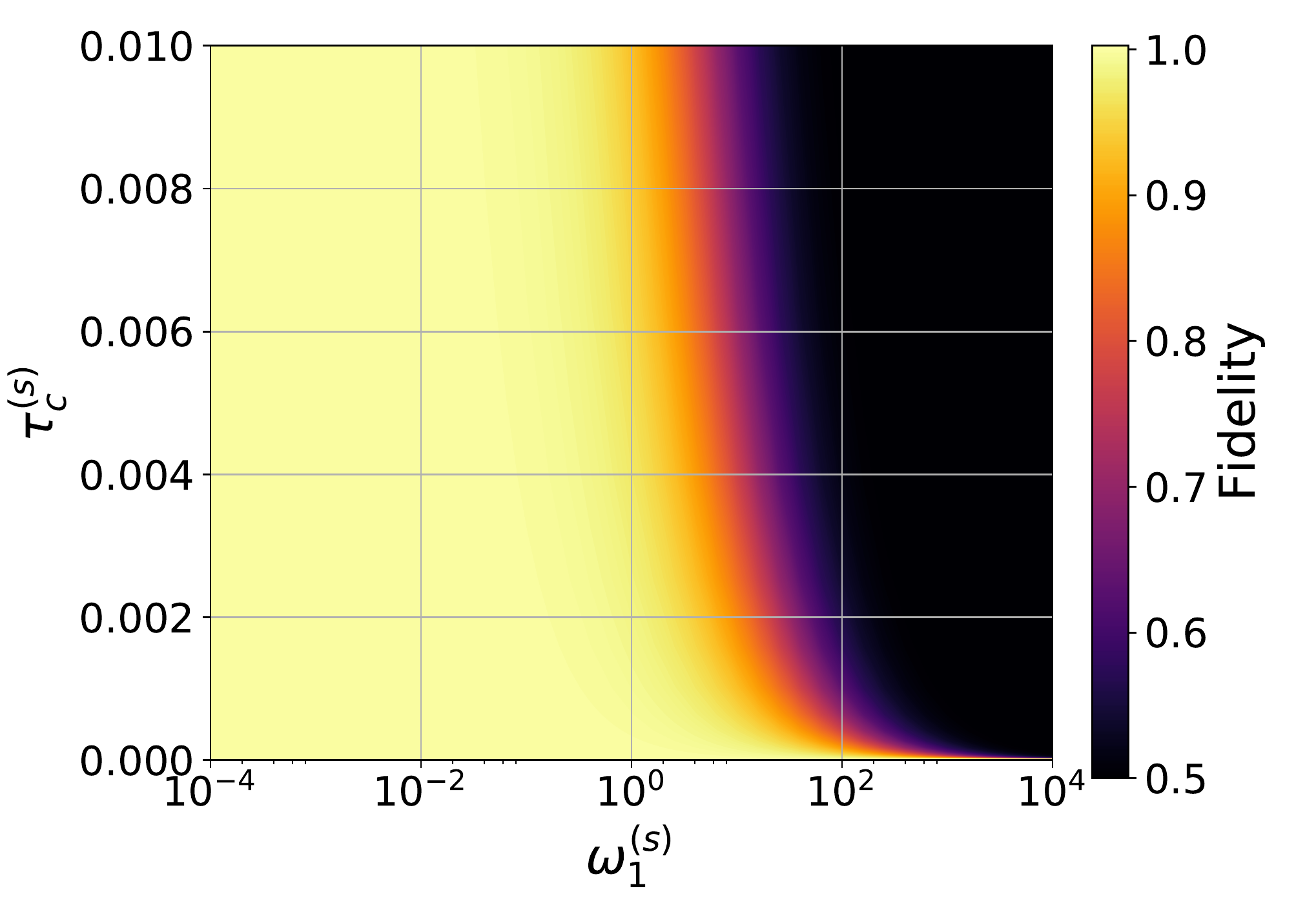} \\
\raisebox{40mm}{(c)}
\includegraphics[width=0.4\linewidth]{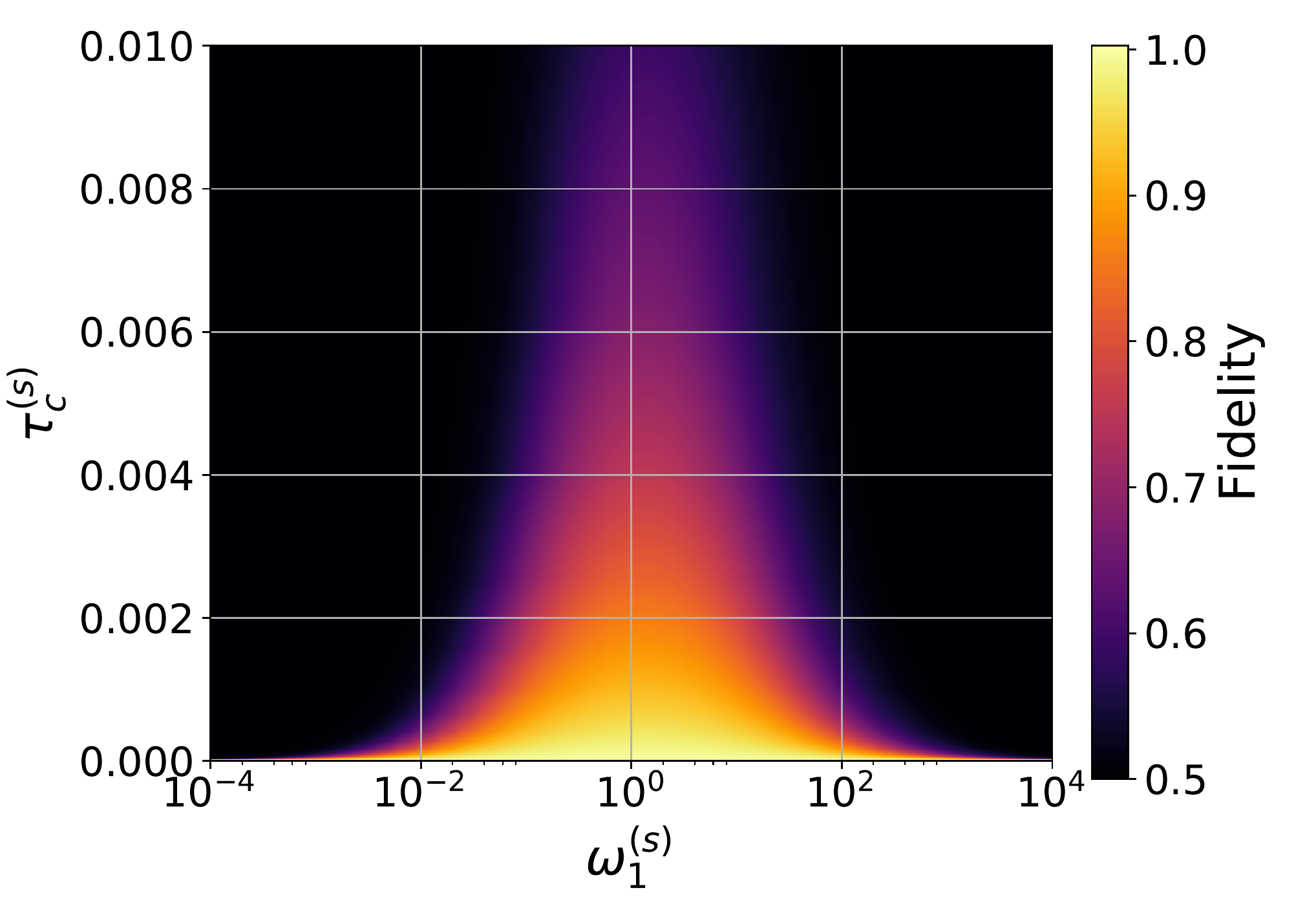}
\caption{Contour plots of fidelity as a function of $\swone$ and $\stauc$, including (a) only relaxation terms,
(b) only DID terms, 
(c) both DID and relaxation terms, for $J = 10$ kHz.}
\label{f2} 
\end{figure}

Then, using (\ref{eq3}) we numerically solve the FRQME (\ref{eqLiouville}) considering both drive and system-environment coupling terms.

In figure \ref{f2}, we have plotted the fidelity of achieving the target state $|\phi_1\rangle$, 
as a function of $\swone$ and $\stauc$, where $\stauc = \wsl\tauc$ and $\swone = \wone/\wsl$ ($\wsl$ being the system-environment coupling strength).
In figure \ref{f2}-(c), we see that there exists an optimum drive amplitude $\wone^{\rm opt} = \wsl$, for which the desired state can be achieved with maximum fidelity. 
Also, we note that the fidelity is relatively high for smaller $\tauc$ values.
As $\tauc \rightarrow 0$ (unitary case), the highest fidelity $F=1$ can be accomplished, as shown by the light yellow (color online; white in print) region at the bottom of the contour plot.

\begin{figure}[h!]
\center
\raisebox{45mm}{(a)}
\includegraphics[width=0.4\linewidth]{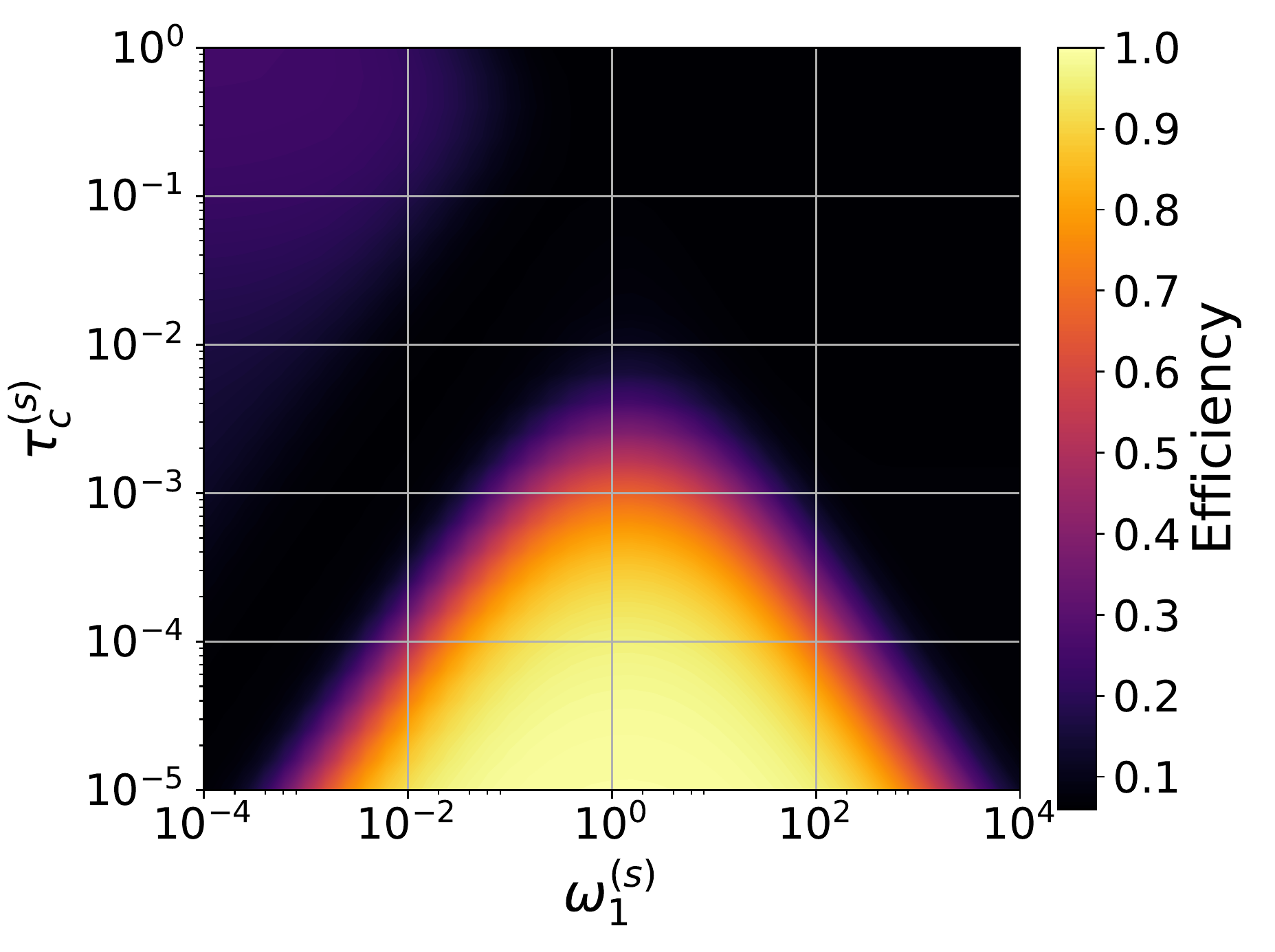}
\raisebox{45mm}{(b)}
\includegraphics[width=0.4\linewidth]{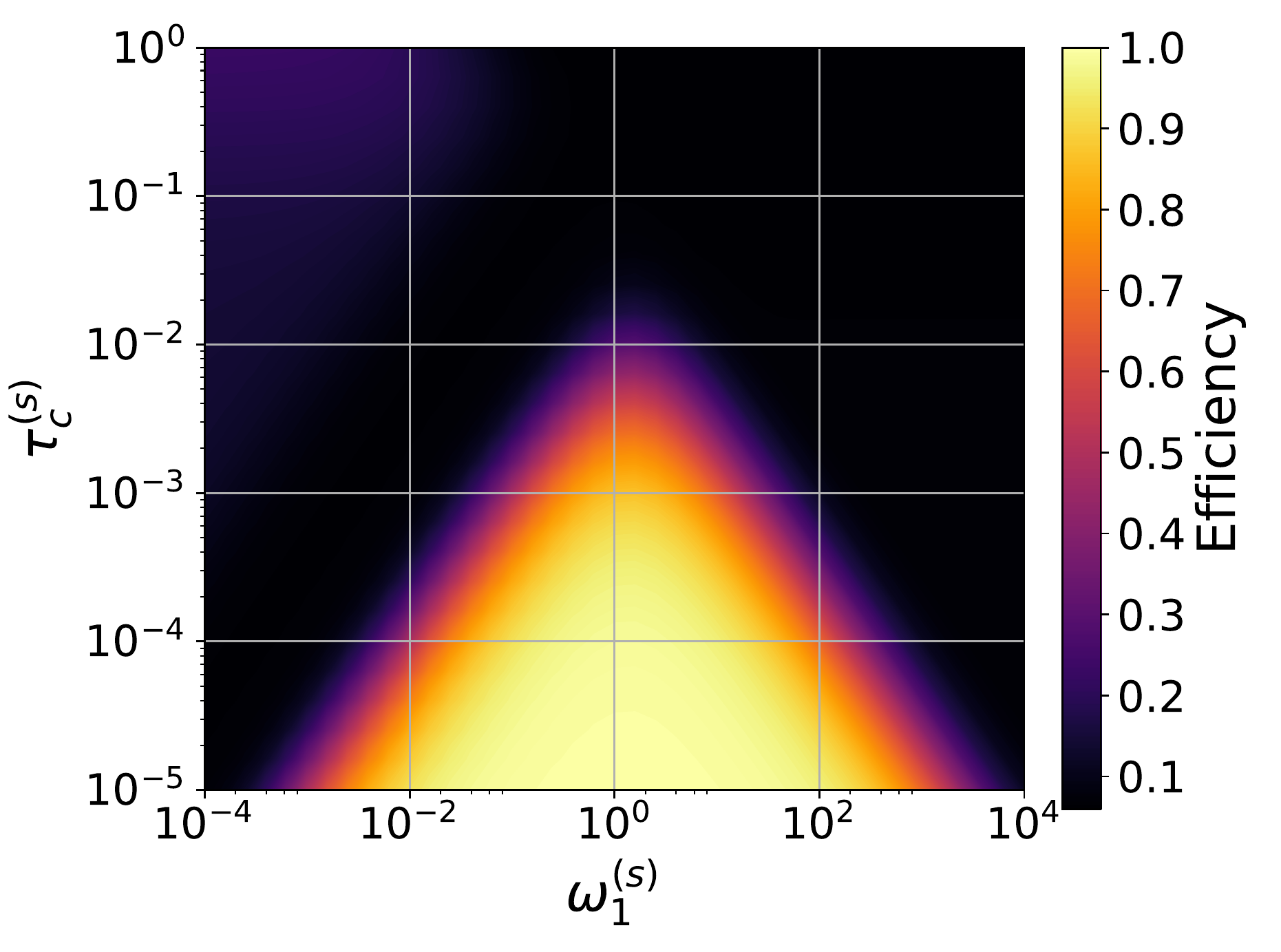}
\caption{Contour plots of efficiency as a function of $\swone$ and $\stauc$, including both DID and relaxation terms for (a) $J = 10$ kHz, (b) $J = 100$ kHz.}
\label{f3} 
\end{figure}

Furthermore, we can quantify the non-unitarity induced in the dynamics by defining another quantity in terms of 
$\mathcal{U}$
only, and call it \textit{efficiency} \cite{chanda_2020}, 
\begin{equation}
\mathcal{E} = {\rm Tr}(\mathcal{U} \mathcal{U}^\dagger)/(2^n)^2
\end{equation} 
$n$ being the number of qubits. 
We plot it for different spin-spin coupling strengths ($J$), as shown in figure \ref{f3}.

\section{Conclusion}

We have shown that the fidelity of the search algorithm exhibits nonmonotonic behavior as a consequence of the decoherence effects arising from two competing processes- DID and system-environment coupling, leading to an optimum drive amplitude ($\wone^{\rm opt}$) for which the fidelity reaches its maximum.
For efficient implementation of the search algorithm on realistic (open quantum system) set-ups, precise knowledge of this drive amplitude would play a crucial role to achieve the maximum possible fidelity.

\end{document}